\documentstyle[preprint,aps]{revtex}
\tightenlines
\begin{document}
\title{Negative high-frequency differential conductivity in semiconductor
superlattices}
\author{Yuriy A. Romanov}
\address{Institute for Physics of Microstructures RAS, 603600 Nizhny Novgorod, \\
Russia}
\author{Lev G. Mourokh and Norman J.M. Horing}
\address{Department of Physics and Engineering Physics, \\
Stevens Institute of Technology, Hoboken, NJ 07030 }
\date{\today }
\maketitle

\begin{abstract}
{We examine the high-frequency differential conductivity response properties
of semiconductor superlattices having various miniband dispersion laws. Our
analysis shows that the anharmonicity of Bloch oscillations (beyond
tight-binding approximation) leads to the occurrence of negative
high-frequency differential conductivity at frequency multiples of the Bloch
frequency. This effect can arise even in regions of positive static
differential conductivity. The influence of strong electron scattering by
optic phonons is analyzed. We propose an optimal superlattice miniband
dispersion law to achieve high-frequency field amplification.}
\end{abstract}


\narrowtext
\newpage

\begin{center}
{\large {\bf I. Introduction} \vspace*{1cm} }
\end{center}

Semiconductor superlattices have been at the focal point of research in the
last few decades because of their unique electronic properties. Their
additional spatial periodicity leads to the formation of narrow Brillouin
minizones ($10^5-10^7 cm^{-1}$) and energy minibands ($10^{-3}-10^{-1}eV$) 
\cite{1,2,3}. Due to the narrowness of these minibands Bloch oscillations 
\cite{4} can be observed to occur even in relatively weak static electric
fields ($10^{2}-10^{4}V/cm$) underscoring the promise of semiconductor
superlattices as a likely mechanism for the amplification of THz signals.
The occurence of Bloch oscillations was confirmed experimentally in a number
of works \cite{BO_exp}.

It is well-known \cite{KSS,K,IR} that within the tight-binding approximation
a superlattice subject to a static electric field with strength $E_c$ can
only amplify fields having frequencies $\omega <\sqrt{\Omega_c^2-\tau^{-2}}$%
, where $\Omega_c=eE_cd/\hbar$, $d$ is the superlattice period, and $\tau$
is the relaxation time. Moreover, amplification is only possible in the
region of current-voltage characteristics having negative static
differential conductivity ($\Omega_c\tau >1$). For $\omega\tau >> 1$ the
real part of high-frequency conductivity, $\sigma (\omega )$, has a
characteristic resonant structure, with its maximal negative value at
frequency $\omega =\Omega_c\sqrt{1-(\Omega_c\tau )^{-2}-2(\Omega_c\tau )^{-1}%
}$, i.e. near $\Omega_c$. The origin of such behavior of the high-frequency
conductivity, associated with the resonant interaction of Bloch oscillations
with the external time-dependent field, as well as with electron bunching in
momentum space, was discussed in Ref.\cite{K} in detail. In the
Wannier-Stark representation \cite{5}, this resonant character is an obvious
consequence of energy conservation in electron transport (transfer along the
Wannier-Stark ladder) accompanied by photon emission/absorption \cite{RR}.

Further consideration beyond the tight-binding approximation, leading to
anharmonicity of Bloch oscillations, can exhibit drastic change in the
character of amplification of time-dependent fields. In particular, the
resonant interaction of Bloch oscillation harmonics with the external field
can give rise to the amplification of harmonic fields with frequencies $%
\omega\sim\nu\Omega_c$, ($\nu$ is an integer). Such amplification can occur
even in regions of current-voltage characteristics having positive static
differential conductivity. This is extremely important in connection with
the necessity of low-frequency domain instability suppression \cite{BO_exp}.

The object of the present paper is the analysis of electron high-frequency
behavior in superlattices having various miniband dispersion laws, and the
exploration of new phenomenology brought about by the anharmonicity of Bloch
oscillations. We compare the following two miniband dispersion laws: \newline
1. The commonly used "sinusoidal" dispersion law in the tight-binding
approximation: 
\begin{equation}
\varepsilon_3(k_3) = {\frac{\Delta}{2}}\left( 1 - \cos(k_3d)\right) ,
\end{equation}
where $\varepsilon_3$ and $k_3$ are electron energy and wave vector along
the superlattice axis, respectively, and $\Delta$ is the miniband width. In
this case there is the region with negative effective electron mass in the
upper half of miniband. \newline
2.The model "parabolic" dispersion law with {\it no} regions of negative
effective mass, but with Bragg reflections at points $k_3=\pm \pi /d$: 
\begin{equation}
\varepsilon_3(k_3) = {\frac{\hbar^2k_3^2}{2m_3}},\hspace*{0.5cm} -{\frac{\pi%
}{d}}<k_3<{\frac{\pi}{d}}.
\end{equation}

This article is structured as follows. In Section II we derive an expression
for high-frequency differential conductivity which is valid for any miniband
dispersion law. The high-frequency electron dynamics for miniband dispersion
laws given by Eq.(1) and Eq.(2) are compared in Section III without
scattering by optic phonons; and in Section IV they are compared in the case
of strong electron-phonon interaction. In Section V we suggest a possible
miniband dispersion law having properties desirable for time-dependent field
amplification and analyze the high-frequency electron dynamics for this
case. Section VI presents the conclusions of our work.

\begin{center}
{\large \vspace*{1cm} {\bf II. General Relations} \vspace*{1cm} }
\end{center}

To analyze the high-frequency differential conductivity, we will determine
the induced superlattice current for an arbitrary miniband dispersion law in
the presence of an external electric field given by 
\begin{equation}
E(t) = E_c + E_1\cos\omega t.
\end{equation}
To start, we employ a Boltzmann equation with a single relaxation time
approximation: 
\begin{equation}
{\frac{\partial f({\bf k},t)}{\partial t}} + {\frac{e{\bf E}(t)}{\hbar}}{%
\frac{\partial f({\bf k},t)}{\partial {\bf k}}} = {\frac{f({\bf k}%
,t)-f_0(\varepsilon ,T)}{\tau}},
\end{equation}
where $f({\bf k},t)$ and $f_0(\varepsilon ,T)$ are the nonequilibrium and
equilibrium distribution functions, respectively, and $T$ is the lattice
temperature. The electric field ${\bf E}$ is applied along the superlattice
axis. Using periodicity in ${\bf k}$-space, we expand functions of interest
in Fourier series: 
\begin{equation}
\varepsilon ({\bf k}) = \sum_{\nu =-\infty}^{\infty}\varepsilon (\nu
,k_{\bot})\exp (i\nu k_3d),
\end{equation}
\begin{equation}
f_0(\varepsilon ,T) = \sum_{\nu =-\infty}^{\infty}F_{\nu}(k_{\bot})\exp
(i\nu k_3d),
\end{equation}
and 
\begin{equation}
f({\bf k},t) = \sum_{\nu =-\infty}^{\infty}F_{\nu}(k_{\bot})\exp (i\nu
k_3d)\Phi_{\nu}(t),
\end{equation}
where $F_{\nu}$ is the Fourier coefficient of the equilibrium distribution
function, 
\begin{equation}
F_{\nu}(k_{\bot}) = {\frac{d}{2\pi}}\int_{-\pi /d}^{\pi /d}f_0(k) \exp
(-i\nu k_3d), \hspace*{0.5cm} F_{\nu}=F_{-\nu}^*,
\end{equation}
and $\Phi_{\nu}$ is the factor by which the Fourier transform of the
nonequilibrium distribution function differs from its equilibrium
counterpart, $F_{\nu}$ ($k_{\bot}$ is the component of electron wave vector
orthogonal to the superlattice axis). Substituting Eq.(7) into Eq.(4), we
obtain the kinetic equation for the multicomponent function $\Phi_{\nu}(t)$
as 
\begin{equation}
\tau {\frac{d\Phi_{\nu}(t)}{dt}}+\left( 1+i\nu\tau\Omega
(t)\right)\Phi_{\nu}(t) = 1, \hspace*{0.5cm} \Omega (t)={\frac{edE(t)}{\hbar}%
}
\end{equation}
with the initial condition 
\begin{equation}
\Phi_{\nu}(0)=1.
\end{equation}
The average superlattice current can be expressed in terms of the functions $%
\Phi_{\nu}(t)$ as 
\begin{equation}
j(t) = {\frac{i}{2}}\sum_{\nu =1}^{\infty}j_{0\nu}\Phi_{\nu}(t) + c.c.,
\end{equation}
where 
\begin{equation}
j_{0\nu} = -{\frac{4ed}{\hbar}}\nu\int {\frac{d^3k}{(2\pi )^3}}%
F_{\nu}(k_{\bot})\varepsilon (k)\exp (i\nu k_3d), \hspace*{0.5cm}
j_{0\nu}^*=-j_{0,-\nu}.
\end{equation}
Solving Eq.(9) with the electric field of Eq.(3) ($eE_1d/\hbar\omega << 1$),
we obtain the complex linear differential superlattice conductivity at
frequency $\omega$ in the presence of a static field $E_c$ as 
\begin{eqnarray}
\sigma (\omega ,\Omega_c) = &&\sum_{\nu =1}^{\infty}{\frac{\sigma_{0\nu}}{%
\left( 1+(\nu\Omega_c\tau )^2-(\omega\tau )^2\right)^2 + 4(\omega\tau )^2}} 
\nonumber \\
&&\times \left( 1-(\nu\Omega_c\tau )^2+(\omega\tau )^2+i\omega\tau {\frac{%
1-3(\nu\Omega_c\tau )^2+(\omega\tau )^2}{1+(\nu\Omega_c\tau )^2}}\right) ,
\end{eqnarray}
where 
\begin{equation}
\sigma_{0\nu} = {\frac{ed\tau}{\hbar}}\nu j_{0\nu}.
\end{equation}
It should be noted that this expression is valid for any miniband dispersion
law (which is involved in Eq.(12) and, consequently, in Eq.(13)).

\begin{center}
{\large \vspace*{1cm} {\bf III. High-frequency electron conductivity} \vspace%
*{1cm} }
\end{center}

In this Section we apply the general miniband results obtained above to the
two miniband dispersion laws given by Eq.(1) and Eq.(2). Substituting Eq.(1)
and Eq.(2) into Eq.(12), we obtain 
\begin{equation}
j_{0\nu} = \delta_{\nu 1}j_0, \hspace*{0.5cm} j_0={\frac{end}{\hbar}}\left( {%
\frac{\Delta}{2}}-\langle\varepsilon_3\rangle_0\right) = {\frac{end\Delta}{%
2\hbar}}I_{\nu}\left( {\frac{\Delta}{2T}}\right) I_0^{-1}\left( {\frac{\Delta%
}{2T}}\right)
\end{equation}
for the "sinusoidal" dispersion law and 
\begin{equation}
j_{0\nu} = (-1)^{\nu +1}{\frac{2\hbar ne}{md\nu}}\exp\left( -{\frac{\pi^2T}{%
4\Delta}}\nu^2\right) , \hspace*{0.5cm} \Delta = {\frac{\hbar^2\pi^2}{2md^2}}
\end{equation}
for the "parabolic" one. Here, $n$ is the electron concentration, $%
\langle\varepsilon_3\rangle_0$ is the average equilibrium longitudinal
electron energy, and $I_{\nu}(x)$ is the modified Bessel function. The
second equality in Eq.(15) is given for nondegenerate Maxwellian statistics
with arbitrary $T$, and Eq.(16) is given for Maxwellian statistics with $%
T<\Delta$.

Accordingly, for superlattices having a "sinusoidal" dispersion law, Bloch
oscillations involve the fundamental harmonic alone, $\sigma_{0\nu}=\sigma_0%
\delta_{1\nu}$ ($\sigma_0$ is the static superlattice conductivity), and
Eq.(13) is in agreement with results of Refs.\cite{KSS,IR}. However, for a
"nonsinusoidal" miniband dispersion law (reflecting the possibility of
electron transport not only to nearby cells) the terms with $\nu >1$ also
contribute to the conductivity (13). Their contribution can be obtained by
replacing of $\Omega_c$ by $\nu\Omega_c$. Formally, this replacement is
equivalent to an increase of the superlattice period by the factor $\nu$,
and, consequently, to a decrease of the Brillouin miniband wavenumber range
by a factor $1/\nu$. To illustrate this behavior, we use the "parabolic"
dispersion law, given by Eq.(2). In this case, according to Eq.(16), we have
for $T\rightarrow 0$ 
\begin{equation}
\sigma_{0\nu}=(-1)^{\nu +1}2\sigma_0.
\end{equation}
Substituting Eq.(17) into Eq.(13) and performing the summation with respect
to $\nu$, we obtain the real part of the high-frequency differential
conductivity as 
\begin{equation}
Re\sigma (\omega ,\Omega_c) = \sigma_0\left( {\frac{1}{1+(\omega\tau )^2}} - 
{\frac{\pi\cosh\left(\pi /\Omega_c\tau\right) \sin\left(\pi\omega
/\Omega_c\right)}{\Omega_c\omega\tau^2\left(\sinh^2\left(\pi
/\Omega_c\tau\right) + \sin^2\left(\pi\omega /\Omega_c\right)\right)}}%
\right) .
\end{equation}
For $\omega >\Omega_c >> \tau^{-1}$, Eq.(18) can be rewritten as 
\begin{equation}
Re\sigma (\omega ,\Omega_c) = -\sigma_0{\frac{\left(\Omega_c/\pi\omega%
\right) \sin\left(\pi\omega /\Omega_c\right)}{1+ \left(\Omega_c\tau
/\pi\right)^2 \sin^2\left(\pi\omega /\Omega_c\right)}}.
\end{equation}
It follows from Eq.(19) that the real part of the high-frequency
differential conductivity changes its sign at $\omega\sim\nu\Omega_c$. For $%
\omega\tau << 1$, Eq.(18) limits to the static differential conductivity
relation 
\begin{equation}
\sigma (0,\Omega_c) = \sigma \left( 1 - \left(\pi /\Omega_c\tau\right)^2{%
\frac{\cosh\left(\pi /\Omega_c\tau\right)}{\sinh^2\left(\pi
/\Omega_c\tau\right)}}\right) ,
\end{equation}
which is negative for $\Omega_c\tau >1.174$. The conductivity of Eq.(18) is
depicted in Figure 1(a) for $\Omega_c\tau = 1.15, 2$, and $5$, respectively.
For comparison, we present the conductivity involving the "sinusoidal"
dispersion law in the Insert. Figure 1(b) clearly shows the resonant
character of the high-frequency superlattice differential conductivity for
the (unrealistic) high value $\Omega_c\tau = 30$. It is evident from the
Figures and from Eq.(19) that the high-frequency superlattice differential
conductivity is negative near the Bloch oscillation harmonics, at
frequencies $\omega < (2\nu +1)\Omega_c$ and $\omega > 2\nu \Omega_c$. This
behavior is a consequence of the sign of the Fourier coefficients of the
"parabolic" dispersion law, as reflected in Eq.(17). One can see that, for
superlattices having a "nonsinusoidal" miniband dispersion law,
amplification of external time-dependent fields is possible for frequencies
larger than the Bloch frequency, near its harmonics and even (most
importantly) in the regions of the current-voltage characteristic having
positive static differential conductivity.

\begin{center}
{\large \vspace*{1cm} {\bf IV. Phonon effects on high-frequency electron
dynamics} \vspace*{1cm} }
\end{center}

In semiconductor structures having strong electron coupling to optic phonons
Bragg reflection may be pre-empted by the emission of a phonon with the
return of the electron to the bottom of the miniband. For simplicity, we
assume that the miniband width, $\Delta$, is equal to or a little larger
than the phonon energy, $\hbar\omega_0$, and that the characteristic time of
phonon emission, $\tau_0$, is shorter than all other relaxation times. If
scattering is absent in the passive region (the region of energies with $%
\varepsilon < \hbar\omega_0$), the electron motion in momentum space is
periodic with frequency $2\Omega_c$, and in coordinate space there are
oscillations with the same frequency as well as a drift with velocity 
\[
\langle V\rangle = {\frac{d}{\pi\hbar}}\int_0^{\pi /d}{\frac{%
\partial\varepsilon (k_3)}{\partial k_3}}dk_3. 
\]
In this case the electron momentum distribution function becomes needle-like
(sharply peaked in the $k_3$-direction, "streaming") and the problem is
really one-dimensional. It should be noted that, in contrast to the case of
a bulk semiconductor, for a superlattice the extent of electron penetration
in the active region (the region of energies with $\varepsilon >
\hbar\omega_0$) is determined not only by the phonon emission time, but by
the miniband width as well. As a result, streaming can be much narrower in
superlattices than in bulk semiconductors. Furthermore, if $\tau_0$ is not
very small, there is a finite probability, ($1-\alpha$), of Bragg reflection
before phonon emission. In this case the electron oscillations have two
characteristic transfer frequencies, $\Omega_c$ and $2\Omega_c$, and the
electron distribution function is again needle-like, but is peaked not only
for positive $k_3$ but also for negative $k_3$. The new cyclic electron
motion with frequency $2\Omega_c$ can result in high-frequency negative
differential conductivity at frequencies close to $2\nu\Omega_c (\nu$ is an
integer). The occurrence of such a negative conductivity in bulk
semiconductors was predicted and studied theoretically in Ref.\cite{AK} and
was confirmed experimentally in Ref.\cite{Vor}. It is a consequence of (a)
electron bunching in momentum space near the involved part of the $k_3$-axis
under the influence of a strong static electric field with optical phonon
scattering, and (b) modulation of the electron momentum distribution by the
additional harmonic field jointly with relaxation processes. The necessary
conditions for such negative conductivity are relatively weak electron
scattering in the passive region ($\Omega_c\tau >1$) and small depth of
penetration in the active region. In summary, streaming in superlattices has
the following features that distinguish it from bulk: 1) the electron
penetration depth in the active region is determined by the miniband width 
{\it independently of the phonon emission time}; 2) streaming occurs for
both positive and negative $k_3$ (double-side streaming) and there are {\it %
two} characteristic transfer frequencies in momentum space; 3) the
dependence of electron velocity on its momentum is {\it nonlinear}.

Electron dynamics in the passive region are described by the Boltzmann
equation (Eq.(4)) with the boundary condition 
\begin{equation}
f\left( -{\frac{\pi }{d}}\right) =(1-\alpha )f\left( {\frac{\pi }{d}}\right)
.
\end{equation}
In the interest of simplicity, we take the initial electron distribution
function in the form 
\begin{equation}
f_{0}(k_{3})=2\pi n\delta (k_{3})
\end{equation}
For convenience, we employ notation with an additional argument inserted
into all functions, the inverse relaxation time, $\tau ^{-1}$. Assuming the
field strength $E_{1}$ in Eq.(3) to be much smaller than the static field
strength $E_{c}$, we write 
\begin{equation}
f(k_{3},t,\tau ^{-1})=f_{c}(k_{3},\tau ^{-1})+f_{1}(k_{3},t,\tau ^{-1}),
\end{equation}
where the static nonequilibrium distribution function $f_c(k_3, \tau^{-1})$
is given by 
\begin{eqnarray}
f_{c}(k_{3}, \tau^{-1})=\frac{2\pi nd}{\Omega _{c}\tau }\frac{\exp \left( -%
\frac{k_{3}d}{\Omega _{c}\tau }\right) }{\left( 1-\exp \left( -\frac{\pi }{%
\Omega _{c}\tau }\right) \right) \left( 1+(1-\alpha )\exp \left( -\frac{\pi 
}{\Omega _{c}\tau }\right) \right) }  \nonumber \\
\times \left\{ 
\begin{array}{c}
1,\hspace*{0.5cm} 0<k_{3}<\pi /d \\ 
(1-\alpha )\exp \left( \frac{2\pi }{\Omega _{c}\tau }\right) ,-\pi
/d<k_{3}<0,
\end{array}
\right.
\end{eqnarray}
and $f_{1}(k_{3},t,\tau ^{-1})=f_{1}(k_{3},\tau ^{-1})\exp (-i\omega t)$
obeys the linearized Boltzmann equation 
\begin{equation}
eE_{c}{\frac{\partial f_{1}(k_{3},\tau^{-1})}{\hbar \partial k_{3}}}+eE_{1}{%
\frac{\partial f_{c}(k_{3},\tau^{-1})}{\hbar \partial k_{3}}}=-(\tau
^{-1}-i\omega )f_{1}(k_{3},\tau^{-1}),
\end{equation}
with the boundary condition of Eq.(21) and the particle conservation
condition 
\begin{equation}
\int_{-\pi /d}^{\pi /d}f_{1}(k_{3},\tau^{-1}){\frac{dk_{3}}{2\pi }}=0.
\end{equation}

The static distribution function of Eq.(24) facilitates the determination of
static current-voltage characteristics. For the "sinusoidal" miniband
dispersion law, it is given by 
\begin{equation}
j_c=\tilde{j}_0{\frac{\Omega_c\tau}{1+(\Omega_c\tau )^2}}{\frac{1-(1-\alpha
)\exp (-\pi /\Omega_c\tau )}{1+(1-\alpha )\exp (-\pi /\Omega_c\tau )}}%
\coth\left( {\frac{\pi}{2\Omega_c\tau}}\right) ,
\end{equation}
whereas in the case of the "parabolic" miniband dispersion law, we have 
\begin{equation}
j_c=\tilde{j}_0\left( {\frac{2}{\pi}}\right)^2\left( \Omega_c\tau - {\frac{%
(2-\alpha )\pi}{2\sinh (\pi /\Omega_c\tau )-\alpha(1-\exp (-\pi
/\Omega_c\tau ))}}\right) .
\end{equation}
Here, $\tilde{j}_0=ne\Delta d/\hbar$. It is important to note that in the
case of high probability of phonon emission, $\alpha\sim 1$, the negative
static differential conductivity is suppressed and can even vanish for any
miniband dispersion law \cite{TBP}. However, in the present paper we are
primarily interested in high-frequency electronic properties of
superlattices and, accordingly, in the distribution function $%
f_1(k_3,t,\tau^{-1})$. The solution of Eq.(25) for this function can be
represented in the form 
\begin{equation}
f_1(k_3,t,\tau^{-1})={\frac{i}{\omega\tau}}{\frac{E_1}{E_c}}%
(f_c(k_3,\tau^{-1})- f_c(k_3,\tau^{-1}-i\omega )).
\end{equation}
Consequently, the high-frequency conductivity is related to static
conductivity (in the single $\tau$-approximation considered here), as given
by 
\begin{equation}
\sigma (\omega ,\tau^{-1})={\frac{i}{\omega\tau}}(\sigma_c(\tau^{-1})-%
\sigma_c(\tau^{-1}-i\omega )).
\end{equation}
It should be emphasized that the closeness of static conductivity values for
different miniband dispersion laws does {\it not} imply closeness of the
corresponding high-frequency conductivities. This is clear from Eq.(30),
since the replacement of relatively small real frequencies ($\tau^{-1}$) by
relatively large complex values $(\tau^{-1}-i\omega )$ leads to the
occurrence of resonant terms and significant phase shifts between current
and applied field.

Only the second terms on the right sides of Eqs.(29,30) contribute to the
real part of the high-frequency conductivity. It is evident from Eqs.(24,29)
that the corresponding part of the distribution function $%
f_1(k_3,t,\tau^{-1})$ contains a modulation factor $\exp\{ i(\omega
dk_3/\Omega_c-\omega t)\}$ describing an electron density wave moving in
momentum space with velocity $eE_c$ and wavelength $\lambda =
2\pi\Omega_c/\omega d$. There is a resonant energy exchange between the
electron density wave and the applied harmonic field occuring at frequencies
close to the transfer frequency harmonics, $\omega\sim 2\nu\Omega_c$ for
single-side streaming, and $\omega\sim (2\nu +1)\Omega_c$ for double-side
streaming. In these cases there is an approximately integer number of wave
lengths $\lambda$ in corresponding transfer regions, $\pi /d$ or $2\pi /d$,
respectively.

The expressions for the high-frequency conductivities are extremely unwieldy
in the general case and we present them only for $\alpha =1$. For the
"sinusoidal" miniband dispersion law, the high-frequency conductivity has
the form 
\begin{equation}
Re\sigma (\omega ) = \sigma_0{\frac{(1+(\omega\tau )^2-(\Omega_c\tau
)^2)\sinh (\pi /\Omega_c\tau )+ (1+(\omega\tau )^2+(\Omega_c\tau )^2)
(\omega\tau )^{-1} \sin (\pi\omega /\Omega_c )}{\left( (1-(\omega\tau
)^2+(\Omega_c\tau )^2)^2+4(\omega\tau )^2\right) (\cosh (\pi /\Omega_c\tau
)-\cos (\pi\omega /\Omega_c))}} ,
\end{equation}
whereas for the "parabolic" miniband dispersion law it is given by 
\begin{equation}
\sigma (\omega ) = \sigma_0\left( {\frac{1+i\omega\tau}{1+(\omega\tau )^2}}-{%
\frac{\pi}{2\Omega_c\omega\tau^2}}{\frac{\sin (\pi\omega /\Omega_c)+2i\coth
(\pi /2\Omega_c\tau )\sin^2(\pi\omega /\Omega_c)}{\cosh (\pi /2\Omega_c\tau
)-\cos (\pi\omega /\Omega_c)}}\right) .
\end{equation}

The real parts of the conductivities of Eqs.(31,32) are presented in Figure
2(a) and Figure 2(b), respectively. In the first case (Fig.2(a)), there is 
{\it no} negative high-frequency differential conductivity at any frequency.
At resonant frequencies, $\omega =2\nu\Omega_c$, the high-frequency
conductivity has maxima which do not depend on $\tau$ for $%
\Omega_c\tau\rightarrow\infty$ (relaxation is only due to phonon emission at
the miniband edge) and are approximately given by 
\begin{equation}
\sigma (\omega )\sim {\frac{2\tilde{j}_0}{\pi (4\nu^2-1)E_c}} .
\end{equation}
In the second case, (Fig.2(b)), the high-frequency conductivity changes sign
as $\omega$ passes through the resonant frequencies, and in their vicinities
(at $\omega\sim 2\nu\Omega_c\pm\tau^{-1}$) it has the form 
\begin{equation}
\sigma (\omega )\sim \mp {\frac{2}{5}}{\frac{\tilde{j}_0}{5\pi^2\nu E_c}},
\end{equation}
which, also, does not depend on $\tau$. The real and imaginary parts of the
high-frequency conductivity for a superlattice calculated from Eqs.(28,30),
are presented in Figures 3(a) and 3(b), respectively, for $\Omega_c\tau =10$
and various values of $\alpha$. Near the even harmonics ($2\nu\Omega_c$)
both real and imaginary parts of the high-frequency conductivity are almost
independent of $\alpha$, because (for "parabolic" case only!) the amplitudes
of even Fourier-harmonics of the electron velocity do not depend on $\alpha$%
. In the vicinity of odd harmonics ($(2\nu+1)\Omega_c$) there is resonant
response only for small $\alpha$. It should be noted that the imaginary part
of the high-frequency conductivity has large negative values which can lead
to plasmon instability in superlattices near the transfer frequencies.

It is of value to understand the reasons for the strong dependence of
high-frequency superlattice conductivity on the miniband dispersion law. The
time-dependent current arises from joint modulation of the electron
distribution function in momentum space by the time-dependent field, Bragg
reflection and scattering. The modulation associated with the time-dependent
field is a result of a homogeneous shift of electrons along the involved
part of the $k_{3}$-axis. For this modulation to be nonzero, it is necessary
that the shifted electron distribution in momentum space (created by the
strong static field and Bragg reflection) should be inhomogeneous. In
particular, without scattering in the passive region and with $\alpha =1$, $%
f_{c}(k_{3})$ is constant and there is no modulation. Furthermore, to induce
large current oscillations, the electron velocity must strongly depend on
momentum. And, finally, to generate negative differential conductivity, the
current phase has to be shifted with respect to the field phase by more than 
$\pi /2$. To examine all these conditions for superlattices, we consider the
region of resonant frequencies, $\omega =2\nu \Omega _{c}+\delta \omega
,|\delta \omega |<<\Omega _{c}$, and strong static fields, i.e. $\Omega
_{c}\tau >>1$. For simplicity we will analyze single-side streaming, so we
assume $\alpha =1$. In this case, according to Eqs.(24,29), the part of the
electron distribution function responsible for the real part of the
high-frequency differential conductivity is given by 
\begin{equation}
\delta f_{1}(k_{3})\approx {\frac{2nE_{1}d}{E_{c}(1+(\tau \delta \omega
)^{2})}}\left[ \tau \delta \omega \sin (2\nu k_{3}d)-\cos (2\nu k_{3}d)%
\right] \equiv \delta f_{1}^{(a)}(k_{3})+\delta f_{1}^{(s)}(k_{3}).
\end{equation}
The first term in the square brackets of Eq.(35), $\delta f_{1}^{(a)}(k_{3})$%
, is antisymmetric with respect to the center of the transfer region (the
point $k_{3}=\pi /2d$), whereas the second term, $\delta f_{1}^{(s)}(k_{3})$%
, is symmetric. We can also rewrite the electron velocity as a sum of
symmetric and antisymmetric parts, $V(k_{3})=V_{a}(k_{3})+V_{s}(k_{3})$. In
this notation, the real part of the high-frequency differential conductivity
has the form 
\begin{eqnarray}
Re\sigma (\omega ) &\approx &{\frac{ned}{\pi E_{c}(1+(\tau \delta \omega
)^{2})}}\left[ \tau \delta \omega \int_{0}^{\pi /d}V_{a}(k_{3})\sin (2\nu
k_{3}d)dk_{3}-\int_{0}^{\pi /d}V_{s}(k_{3})\cos ((2\nu +1)k_{3}d)\right]  
\nonumber \\
&\equiv &\sigma _{a}(\omega )+\sigma _{s}(\omega ).
\end{eqnarray}
The conductivity $\sigma _{a}(\omega )$, determined by the antisymmetric
velocity component, changes sign as $\omega $ passes through the resonant
frequencies, $2\nu \Omega _{c}$, whereas the conductivity $\sigma
_{s}(\omega )$, determined by the symmetric velocity component, does not
change sign, and it is positive, as usual (when $V_{s}(k_{3})$ increases
with increasing $k_{3}$ near the bottom of the miniband and decreases near
the top). For superlattices having the ''sinusoidal'' miniband dispersion
law, $V(k_{3})=V_{s}(k_{3})$, and for the superlattices having the
''parabolic'' miniband dispersion law, $V(k_{3})=V_{a}(k_{3})+const$. As a
result, negative high-frequency differential conductivity occurs in the
latter case and does not exist in the former case.

\begin{center}
{\large \vspace*{1cm} {\bf V. Optimal miniband dispersion law for
high-frequency field amplification} \vspace*{1cm} }
\end{center}

In the preceding Section we showed that the value of the negative
high-frequency differential conductivity at the transfer frequencies depends
not only on the electron scattering mechanisms in the passive region and the
depth of penetration into the active region, but it is also very sensitive
to the miniband dispersion law. In particular, contrary to the usual
expectation, the existence of a negative effective electron mass region
actually militates against the occurrence of negative high-frequency
conductivity. Minibands having much higher electron velocities in the top
part of the band than those in the bottom part are optimal for negative
high-frequency differential conductivity. In this case modulation of the
electron distribution function in momentum space induces a large
time-dependent current, and, moreover, the contribution of the symmetric
velocity component to the high-frequency differential conductivity becomes
negative at the resonant frequencies. To explore this, we consider the
conductivity of a structure having a "superquadratic" miniband dispersion
law, as given by 
\begin{equation}
\varepsilon (k_3)={\frac{\hbar^2}{2}}\left\{ 
\begin{array}{c}
{\frac{k_3^2}{m_1}},\hspace*{0.5cm} 0<\vert k_{3}\vert <{\frac{\pi}{2d}} \\ 
{\frac{k_3^2}{m_2}}-{\frac{\pi}{d}}\left( {\frac{1}{m_2}}-{\frac{1}{m_1}}%
\right) \left( \vert k_3\vert -{\frac{\pi}{4d}}\right) ,{\frac{\pi}{2d}}%
<\vert k_{3}\vert <{\frac{\pi}{d}}
\end{array}
\right. .
\end{equation}
In the region $0<\vert k_{3}\vert <\pi /2d$, electrons have a positive
effective mass $m_1$ and in the region $\pi /2d<\vert k_{3}\vert <\pi /d$,
they have a different positive effective mass $m_2$. Qualitatively similar
dispersion laws are found in hole quantum layers \cite{Bekin}. For such a
dispersion law, the symmetric and antisymmetric electron velocity components
are given by (for $k_3>0$) 
\begin{equation}
V_a(k_3)={\frac{\hbar}{2}}\left( {\frac{1}{m_2}}+{\frac{1}{m_1}}\right)
\left( k_3 -{\frac{\pi}{2d}}\right) ,
\end{equation}
and 
\begin{equation}
V_s(k_3)={\frac{\hbar}{2}}\left( {\frac{1}{m_2}}-{\frac{1}{m_1}}\right)
\left\vert k_3 -{\frac{\pi}{2d}}\right\vert +{\frac{\hbar\pi}{2m_1d}}.
\end{equation}
Substituting Eqs.(38,39) into Eq.(36), we obtain 
\begin{equation}
Re\sigma (\omega ) \approx {\frac{-\sigma_0}{4\nu (1+(\tau\delta\omega )^2)}}%
\left( (1+\eta ){\frac{\delta\omega}{\Omega_c}}+{\frac{1-(-1)^{\nu}}{\pi\nu}}%
{\frac{\eta -1}{\Omega_c\tau}}\right) ,
\end{equation}
where $\eta =m_1/m_2$. It follows from Eq.(40) that, for $\eta >1$, the
contribution of the symmetric velocity component to the high-frequency
conductivity is negative, and, for $\eta <1$, it is positive. The case with $%
\eta =-1$ corresponds to the "sinusoidal" dispersion law. The real part of
the high-frequency differential conductivity is presented in Figure 4 for
various ratios of effective masses, and with $\Omega_c\tau =3$. It should be
noted that negative high-frequency conductivity occurs even for $\eta <1$,
but its magnitude is much smaller than that for $\eta >1$. For very large $%
\eta$, negative high-frequency conductivity arises even for small static
fields ($\Omega_c\tau <1$), when the low-frequency differential conductivity
is large and positive. To illustrate this, we present the high-frequency
differential conductivity of a superlattice having the miniband dispersion
law of Eq.(37) for $\eta =10, 30,$ and $50$ and $\Omega_c\tau=0.8$.

In the analysis above we did not take into consideration the field
dependence of the probability of phonon emission, $\alpha$. Account of this
dependence would lead to an additional modulation of the electron
distribution in momentum space and some modification (but not vanishing!) of
the range of negative high-frequency differential conductivity. It should be
noted that interminiband tunneling could also give rise to the formation of
negative high-frequency differential conductivity in the regions having
positive static differential conductivity \cite{OR}.

\begin{center}
{\large \vspace*{1cm} {\bf VI. Conclusions} \vspace*{1cm} }
\end{center}

In summary, our analysis shows that the anharmonicity of Bloch oscillations
in superlattices having a "nonsinusoidal" miniband dispersion law (beyond
the tight-binding approximation) should lead to amplification of harmonic
fields with frequencies that are multiples of the Bloch frequency. Such
amplification can occur even in regions of the current-voltage
characteristics that have a positive static differential conductivity. We
have compared high-frequency electron behavior in superlattices having (i)
"sinusoidal" (Eq.(1)) and (ii) "parabolic" (Eq.(2)) miniband dispersion laws
for the cases (a) without optic phonon scattering and (b) with strong
electron-optic phonon coupling. We have obtained explicit expressions for
the high-frequency differential conductivity in all these situations
exhibiting the regions where it is negative. On this basis, we propose the
"superquadratic" miniband dispersion law (Eq.(37)) as optimal for
high-frequency field amplification. This dispersion law can be realized in a
superlattice of quasi-two-dimensional hole layers.

\begin{center}
\vspace*{1cm}

{\large {\bf Acknowledgements} }

\vspace*{1cm}
\end{center}

The work of Yu.A.R. is supported by the Russian Foundation for Basis
Research (Grant No. 01-02-16446) and by the Program "Low-Dimensional Quantum
Structures" of Russian Academy of Science, L.G.M. and N.J.M.H. gratefully
acknowledge support from the US Department of Defense, DAAD 19-01-1-0592.

\begin{figure}[tbp]
\caption{High-frequency differential conductivity for the "parabolic"
miniband dispersion law (Eq.(2)); (a) for $\Omega_c\protect\tau =1.15;2;5$,
(b) for $\Omega_c\protect\tau =30$. Inserts: high-frequency conductivity of
the "sinusoidal" miniband dispersion law (Eq.(1)).}
\label{fig1}
\end{figure}

\begin{figure}[tbp]
\caption{High-frequency differential conductivity for single-side streaming
with $\Omega_c\protect\tau =3;5;10$; (a) for the "sinusoidal" miniband
dispersion law; (b) for the "parabolic" miniband dispersion law.}
\label{fig2}
\end{figure}

\begin{figure}[tbp]
\caption{High-frequency differential conductivity for double-side streaming
for the "parabolic" miniband dispersion law with $\Omega_c\protect\tau =10$
and $\protect\alpha =0;0.5;1$; (a) real part; (b)imaginary part.}
\label{fig3}
\end{figure}

\begin{figure}[tbp]
\caption{High-frequency differential conductivity for the "superquadratic"
miniband dispersion law (Eq.(37)) with $\Omega_c\protect\tau =3$ and $%
\protect\eta =1;5;10$.}
\label{fig4}
\end{figure}

\begin{figure}[tbp]
\caption{High-frequency differential conductivity for the "superquadratic"
miniband dispersion law with $\Omega_c\protect\tau =0.8$ and $\protect\eta %
=10;30;50$.}
\label{fig5}
\end{figure}

\end{document}